\documentstyle[amssymb,preprint,aps]{revtex}
\tightenlines 
\begin{document}
\draft
\title{Two harmonically coupled Brownian particles in random media\ }
\author{M. Schulz, S. Stepanow, S. Trimper}
\address{Martin-Luther-Universit\"{a}t Halle-Wittenberg, Fachbereich Physik, D-06099\\
Halle, Germany}
\date{\today }
\maketitle

\begin{abstract}
We study the behaviour of two Brownian particles coupled by an elastic
harmonic force in a quenched disordered medium. We found that to first order
in disorder strength, the relative motion weakens (with respect to the
reference state of a Brownian particle with the double mass) the effect of
the quenched forces on the centre of mass motion of the Brownian particles,
so that the motion will become less subdiffusive (superdiffusive) for
potential (solenoidal) disorder. The mean-square relative distance between
the particles behaves in a different way depending of whether the particles
are free to move or one particle is anchored in the space. While the effect
of nonpotential disorder consists in increasing the mean-square distance in
both cases, the potential disorder decreases the mean-square distance, when
the particles are free to move, and increases it when one particle is
anchored in the space.
\end{abstract}


Diffusion of a particle in quenched random environment has been the subject
of numerous studies in the past decade \cite{sinai}-\cite{bouchaud}. It is
of interest in problems like tracer diffusion in porous media, transport in
nonsymmetrical hopping systems, diffusion in a fluid with stationary random
streams. For short range disorder the diffusion is anomalous, $%
<x^{2}(t)>\,\sim t^{2\nu }$ with $\nu \neq 1/2$, for space dimensions lower
than $2$. In this $\ $Letter we present the results of the study of the
behaviour of two Brownian particles inserted in the same disordered media
and coupled by a harmonic force. The main interest in this problem is due to
the problem of the behaviour of a polymer chain in disordered media \cite
{muthukumar}-\cite{goldschmidt}. Two harmonically coupled Brownian particles
in disordered media may be considered as the simplest case of a polymer
chain, which has only two modes: the centre of mass mode and the relative
mode, and thus may serve as a toy model of the much more complicated problem
of a polymer in disordered media. We will study in the present Letter the
interplay between the centre of mass motion and the relative motion in the
presence of quenched disorder. Recently, the interplay between the centre of
mass motion and the relative motion has been considered in the case of two
electrons in disordered media \cite{dorokhov90}-\cite{vonoppen}. The task we
study here is the classical counterpart of the problem of two coupled
electrons in disordered media \cite{dorokhov90}- \cite{vonoppen}.

The Langevin equations for the coupled Brownian particles are 
\begin{equation}
\gamma \frac{dx_{1}}{dt}+\lambda (x_{1}-x_{2})=\eta _{1}(t)+F(x_{1}),\hspace{%
1cm}\gamma \frac{dx_{2}}{dt}+\lambda (x_{2}-x_{1})=\eta _{2}(t)+F(x_{2}),
\label{bt1}
\end{equation}
where $\gamma $ is the monomer friction coefficient, $\lambda $ is the
spring constant, $\eta _{1}(t)$ is the thermal noise, which is assumed to be
Gaussian distributed with zero mean and the correlator $\overline{\eta
_{1}^{i}(t)\eta _{1}^{j}(t^{\prime })}=2\gamma kT\delta ^{ij}\delta
(t-t^{\prime })$ and $kT/\gamma =D_{0}$ being the diffusion constant of the
Brownian particle. The quenched random force $F(x)$ is assumed to be
Gaussian distributed with the zero mean and the correlator $%
<F^{i}(x)F^{j}(x^{\prime })>\,=C^{ij}(x-x^{\prime })$, where the Fourier
transform of $C^{ij}(x-x^{\prime })$ is given by 
\begin{equation}
C^{ij}(q)=q^{-2\varkappa }(Aq^{i}q^{j}/q^{2}+B(\delta
^{ij}-q^{i}q^{j}/q^{2})).  \label{dis}
\end{equation}

By using the Martin-Siggia formalism \cite{msr} the Langevin equations (\ref
{bt1}) can be represented as a path integral over the trajectories of the
Brownian particles in the phase space, which is composed of the coordinates
of the particles, $x_{\alpha }(t)$, and the response fields (momenta), $\
p_{\alpha }(t)$, with the weight determined by the action $S=S_{0}+S_{i}$
with $S_{0}$ and $S_{i}$ given by 
\begin{eqnarray}
&&S_{0}=kT\gamma \int dt^{\prime }p_{\alpha }^{2}(t^{\prime })-i\int
dt^{\prime }p_{\alpha }(t^{\prime })(\gamma dx_{\alpha }(t^{\prime
})/dt^{\prime }-i\lambda \int dt^{\prime }p_{1}(t^{\prime })(x_{1}(t^{\prime
})-x_{2}(t^{\prime }))-  \nonumber \\
&&i\lambda \int dt^{\prime }p_{2}(t^{\prime })(x_{2}(t^{\prime
})-x_{1}(t^{\prime }))),  \label{bt2}
\end{eqnarray}
\begin{equation}
S_{i}=\frac{1}{2}\int dt_{1}\int dt_{2}^{\prime }p_{\alpha }(t_{1})C_{\alpha
\beta }(x_{\alpha }(t_{1})-x_{\beta }(t_{2}))p_{\beta }(t_{2}),  \label{bt3}
\end{equation}
where the sum convention over Greek indices is used above.

To study the perturbation expansions of the quantities such as the centre of
mass and the relative distance between the particles we consider the
correlation function 
\begin{equation}
<\exp (p_{c}x_{c}(t)+ip_{r}x_{r}(t))>_{0},  \label{bt4}
\end{equation}
where the coordinate of the centre of mass $x_{c}$ and the relative
coordinate $x_{r}$ are defined as follows $x_{c}=(x_{1}+x_{2})/2$, and $%
x_{r}=x_{1}-x_{2}$. This transformation of the coordinates implies the
transformation of the response fields in (\ref{bt2}) and momenta in (\ref
{bt4}) as follows $p_{c}=p_{1}+p_{2}$ and $p_{r}=(p_{1}-p_{2})/2$. The
subscript $0$ in Eq.(\ref{bt4}) means that $x_{c}(t_{0})$ and $x_{r}(t_{0})$
are fixed at the initial time $t_{0}$. The average over $x_{r}(t_{0})$ with
the equilibrium Boltzmann weight is obtained by taking the limit $%
t_{0}\rightarrow -\infty $. In studying $<x_{c}(t)^{2}>$ we consider $%
x_{c}(t_{0})$ to be fixed ($=0$), while in computing $<x_{r}(t)^{2}>$ we
proceed as follows. We confine the centre of mass mode in a harmonic
potential with an elastic constant $\lambda _{0}$, let $t_{0}$ tend to $%
-\infty $ and then put $\lambda _{0}=0$. This procedure corresponds to the
integration over $x_{c}(t_{0})$ in volume $\sim (D_{0}/\lambda _{0})^{d/2}$.
Due to the procedures of carrying out the averages over the initial
conditions the results we obtain\ below, Eqs.(\ref{bt5}, \ref{bt9}), are
restricted from below by a small nonzero value of the harmonic coupling $%
\lambda $.

The average in Eq.(\ref{bt4}) can be represented as path integral over the
phase trajectories of Brownian particles weighted with MSR functional (\ref
{bt2}-\ref{bt3}). Due to introduction of the centre of mass and relative
coordinates the free part of the action $S_{0}$ splits respectively into two
independent parts dependent of the centre of mass and the relative
coordinates, which, however, are coupled due to the interaction part of the
action $S_{i}$.

The mean-square displacement $<x_{c}(t)^{2}>$ of the centre of mass is
obtained by using (\ref{bt4}) to first order in powers of the disorder
strength as 
\begin{eqnarray}
&<&x_{c}(t)^{2}>\,=2dD_{c}t+\frac{S_{d}}{2(2\pi )^{d}}\Gamma (d/2-\varkappa
)D_{c}^{-d/2+\varkappa }\int_{0}^{t}dy(t-y)\times  \nonumber \\
&&\{(A+dB-B)(T(y)^{-d/2+\varkappa }+U(y)^{-d/2+\varkappa
})-2Ay(d/2-\varkappa )\times  \nonumber \\
&&(T(y)^{-1-d/2+\varkappa }+U(y)^{-1-d/2+\varkappa }-\exp (-\overline{%
\lambda }y)(U(y)^{-1-d/2+\varkappa }-T(y)^{-1-d/2+\varkappa }\},  \label{bt5}
\end{eqnarray}
where $T(y)=y+(1-\exp (-\overline{\lambda }y))/(\overline{\lambda })$, $%
U(y)=y+(1+\exp (-\overline{\lambda }y))/(\overline{\lambda })$ with $%
\overline{\lambda }=2\lambda /\gamma $, $D_{c}=D_{0}/2$. In the limit of
large $\lambda $ we obtain from Eq.(\ref{bt5}) the first-order correction to
the mean-square displacement of a Brownian particle with the diffusion
constant $D_{c}$. After carrying out the integration over time we obtain 
\begin{eqnarray}
&<&x(t)^{2}>\,=2dD_{c}t(1+\frac{1}{2d}\frac{S_{d}}{2(2\pi )^{d}}\Gamma
(d/2-\varkappa )D_{c}^{-2}2(A+dB-B-2A(d/2-\varkappa ))\times  \nonumber \\
&&\frac{1}{(1+\varkappa -d/2)(2+\varkappa -d/2)}(D_{c}t)^{1+\varkappa
-d/2}+...).  \label{bt6}
\end{eqnarray}
We have found that the preasymptotic term in the expansion of Eq.(\ref{bt5})
for large $\lambda $ does not possess the $1/\varepsilon $-pole ($%
\varepsilon =d_{c}-d$) in the vicinity of the upper critical dimension $%
d_{c}=2+2\varkappa $. Due to the nonexistence of the fixed-point for the
effective coupling constant for potential disorder quantitative results
beyond the first-order perturbation theory are hard to be made.

In Fig.1 we present the results of the numerical evaluation of the integral
in Eq.(\ref{bt5}) for dimensionalities $d=1.5$ and $d=1$. For potential
field the correction decreases in both cases with increasing $\lambda $.
This can be explained qualitatively as follows. When one particle is under
the action of a well, its partner, which is outside the well will diminish
the effect of the well, with the result that the first order correction will
become less subdiffusive. The qualitative run of lines of the force is shown
in Fig.2a. The total effect of the disorder on the behavior of the particle
is hardly to be predicted, because for potential disorder the effective
coupling increases under renormalization \cite{luck}-\cite{honkonen}. Fig.1b
shows the dependence on $\lambda $ for $d=1$. The first-order correction is
positive and decreases to zero with increase of $\lambda $, in agreement
with the fact that the first-order correction is zero for one Brownian
particle \cite{luck}-\cite{honkonen}. In this case the partner contribute to
average out the effect of the barriers and wells acting both subdiffusive,
so that for finite $\lambda $ the correction is positive (superdiffusive). A
prediction of the whole effect of the disorder cannot be made by the present
study.

The reason for superdiffusion in the solenoidal field is due to the remnant
of the ballistic motion of the particle along the lines of the force (see
Fig.2b), which is disturbed by the thermal noise. For two particles the
effect is expected to be maximal, if both particles follows the same line of
the force, which takes place, if the particles are tied closely together
(large $\lambda $). Thus, for intermediate $\lambda $ the superdiffusive
effect of the solenoidal field will be reduced in agreement with the
behaviour shown in Fig.1a.

We now will consider the relative motion of two Brownian particles in the
case when one of the particles is anchored in the space. This problem is
equivalent to the case of one particle subjected to the external harmonic
potential and quenched random forces. The mean-square distance between the
particles is obtained as 
\begin{eqnarray}
&<&x_{r}^{2}>\,=\frac{2dD_{0}}{2\lambda }+\frac{S_{d}\Gamma (d/2-\varkappa )%
}{2(2\pi )^{d}\lambda }D_{0}^{-d/2+\varkappa }\int_{0}^{\infty }dy\exp
(-\lambda y)\{(A+dB-B)T_{1}(y)^{-d/2+\varkappa }  \nonumber \\
&&-\frac{A(d/2-\varkappa )}{\lambda }(1-\exp (-\lambda
y))T_{1}(y)^{-1-d/2+\varkappa }\},  \label{bt7}
\end{eqnarray}
where $T_{1}(y)=(1-\exp (-\lambda y))/\lambda $ and $\lambda $ in (\ref{bt7}-%
\ref{bt8}) is measured in units of $1/\gamma $. Carrying out the
integrations over $y$ and $q$ yields 
\begin{equation}
<x_{r}^{2}>\,=\frac{2dD_{0}}{2\lambda }(1+\frac{S_{d}\Gamma (d/2-\varkappa )%
}{2d(2\pi )^{d}D_{0}^{2}}(A+\frac{(d-1)B}{1-d/2+\varkappa })(\frac{D_{0}}{%
\lambda })^{1-d/2+\varkappa }+...).  \label{bt8}
\end{equation}
Eq.(\ref{bt8}) shows that the first-order correction is positive for
disorder given by (\ref{dis}), so that the effect of disorder consists in
increasing the distance between the particles. The increase of the relative
distance between the particles can be interpreted as the decrease of the
elastic constant. In the case $A=B$ and $\varkappa =0$ this result was
obtained previously in \cite{stepanow/kraft91}. The increase of \ the
mean-square distance between the particles can be understood qualitatively
as follows. The interplay between the harmonic potential and the quenched
disorder becomes important on distances where the harmonic potential is
comparable to $kT$, $r^{2}\simeq D_{0}/\lambda $. The distances with $r>%
\sqrt{D_{0}/\lambda }$, which are practically unaccessible in the absence of
disorder, may become accessible, if the quenched random forces compensate
the harmonic potential. Due to this the mean-square distance from the origin
or the relative distance between the particles increases. Notice that
according to Eq.(\ref{bt8}) the first-order correction for potential
disorder ($B=0$) is regular at the critical dimension, $1-d_{c}/2+\varkappa
=0$, while for solenoidal disorder ($A=0$) the first-order correction
diverges at the critical dimension $d_{c}$.

We now will consider the first-order correction to the mean-square distance
between the Brownian particles in the limit $t\rightarrow \infty $ by taking
into account the diffusion of the centre of mass. In computing $%
\lim_{t\rightarrow \infty }<x_{r}^{2}(t)>$ we confine the centre of mass
mode in a harmonic potential with a small elastic constant $\lambda _{0}$,
carry out the limit $t_{0}\rightarrow -\infty $ and then put $\lambda _{0}=0$%
. This procedure corresponds to the integration of the centre of mass over
the volume $(D_{0}/\lambda _{0})^{d/2}$. As a result we have obtained 
\begin{eqnarray}
&<&x_{r}^{2}>\,=\frac{2dD_{0}}{2\overline{\lambda }}(1+\frac{S_{d}\Gamma
(d/2-\varkappa )}{2d(2\pi )^{d}(D_{0}/2)^{2}}(D_{0}/\overline{\lambda }%
)^{1-d/2+\varkappa }\{-A+  \nonumber \\
&&2^{d/2-\varkappa -1}(d-1)B\int_{0}^{\infty }dy\exp (-y)(T_{\overline{%
\lambda }=1}(y)^{-d/2+\varkappa }-U_{\overline{\lambda }=1}(y)^{-d/2+%
\varkappa })\}+...).  \label{bt9}
\end{eqnarray}
The integral over $y$ in Eq.(\ref{bt9}) can be tabulated for particular
values of $d$ and $\varkappa $. For $d=1$ and $\varkappa =0$ it is equal to $%
0.669$. It increases with increasing $d$. It follows from Eq.(\ref{bt9})
that the potential disorder ($B=0$) results in a decrease of the mean-square
relative distance between the particles. As it follows from Eq.(\ref{bt8})
this is in contrast to the case when one particle is anchored in the space.

Let us discuss now the predictions of Eq.(\ref{bt9}) qualitatively. For
solenoidal field the correction to $<x_{r}^{2}>$ is positive for both cases
with and without the motion of the centre of mass. The solenoidal field,
which does not possess sources and sinks acts rather as an additional
thermal noise, which consequently will increase the distance between the
particles. In the case of the potential disorder the effect with moving
centre of mass is opposite to that discussed above when one particle is
anchored. The decrease of the relative distance can be explained to be due
to the fact that both particles will prefer to be in the regions where the
potential has low values. A similar behaviour was predicted by Cates and
Ball \cite{cates} to occur for a polymer chain in disordered media. The
decrease of the effect of the wells on the centre of mass motion due to the
harmonic coupling between the particles is consistent with the decrease of
the relative distance. Notice that the correction is regular at the upper
critical dimension, so that at this order the effect is small, and is not
expected to give a rise for a power law for the effective elastic constant.
The decrease of the distance is in agreement with the prediction made in 
\cite{ledoussal} for polymer chain, however these authors do not state the
difference between potential and solenoidal disorder. To our knowledge the
present work is the first one where the difference in the behaviour with and
without the motion of the centre of mass is demonstrated in a
straightforward calculation. The first-order correction to the mean-square
relative distance is however regular at the critical dimension for both
cases associated with Eqs.(\ref{bt8}) and (\ref{bt9}). The first-order
correction to $<x_{r}^{2}>$ for solenoidal disorder ($A=0$) is positive and
contains a $1/\varepsilon $-pole in the vicinity of the critical dimension.

To conclude, we have considered the behaviour of two Brownian particles
coupled by an harmonic potential in a disordered medium to first order of
the perturbation theory. We have found that the harmonic coupling weakens
(with respect to the reference state of a Brownian particle with the double
mass) the effect of the quenched forces on the centre of mass motion of the
Brownian particles, i.e. the relative motion contribute in averaging out the
effect of the disorder. For solenoidal (potential) disorder, the centre of
mass motion will become less superdiffusive (subdiffusive). The mean-square
relative distance between the particles behaves in a different way depending
on whether the Brownian particles are free to move or one particle is
anchored in the space. While the effect of solenoidal disorder is the same
in both cases, and consists in increasing the mean-square distance, the
potential disorder decreases the mean square distance, when the particles
are free to move, and increases it when one particle is anchored in the
space.

\acknowledgments A support from the Deutsche Forschungsgemeinschaft (SFB
418) is gratefully acknowledged. S.S. acknowledges also the financial
support from DFG, grant Ste-981/1-1.

\newpage Figure captions

Fig.1 Dependence of the correction to $<x_{c}(t)^{2}>$ as function of $%
\lambda $. a) $d=1.5$, $\kappa =0$; b) $d=1$, $\kappa =0$.

Fig.2 The lines of force of a) the potential and b) solenoidal quenched
random field.

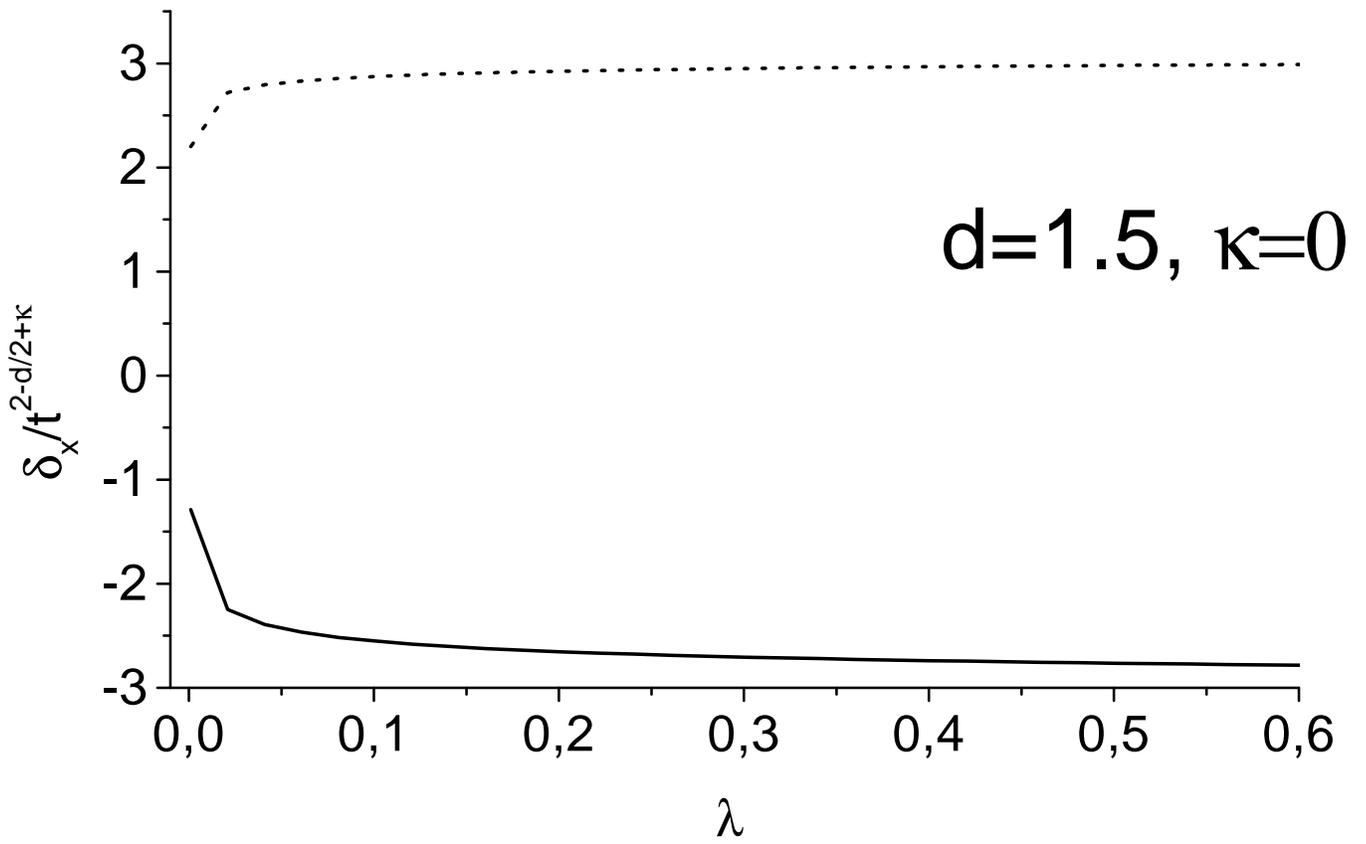
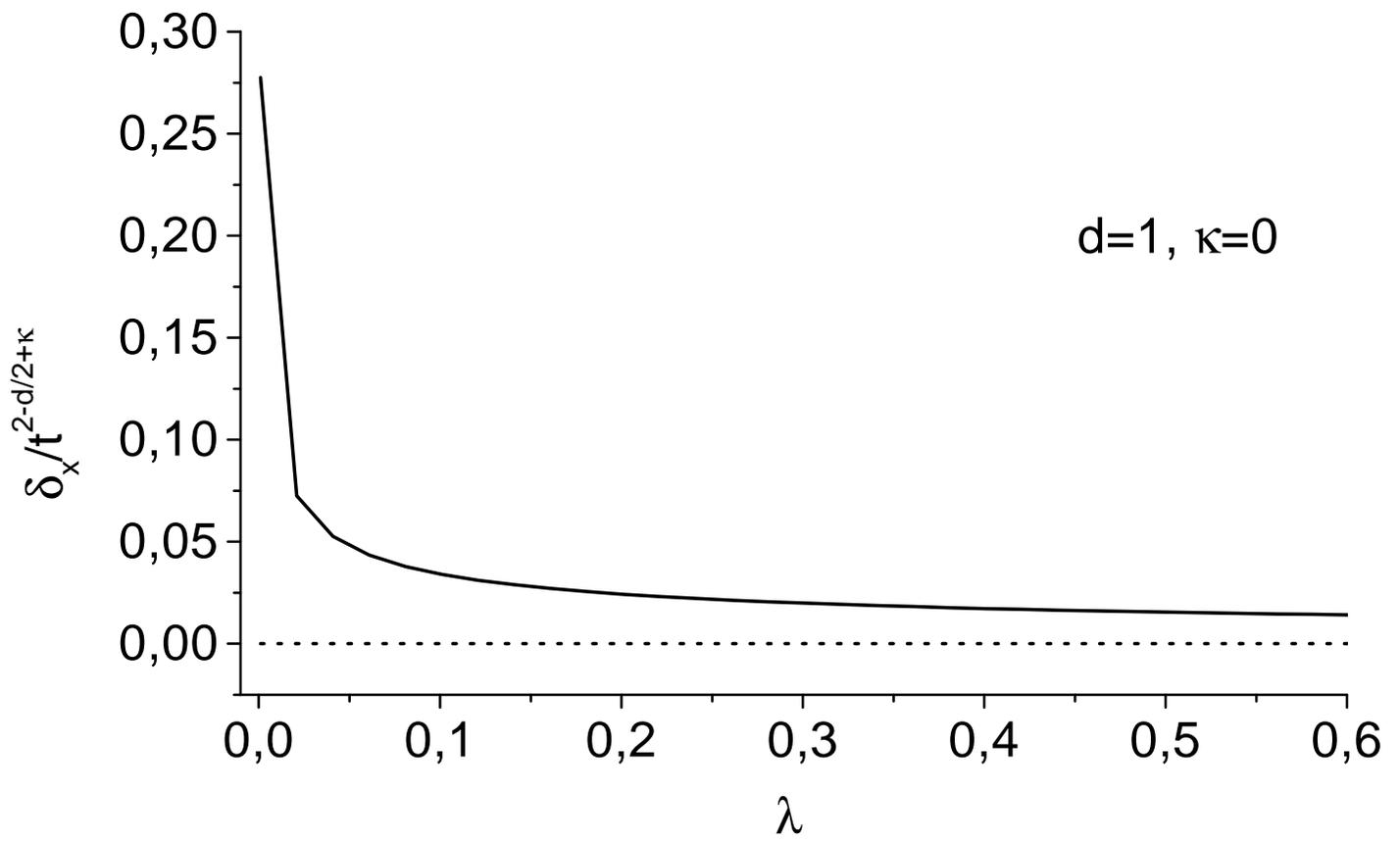
\end{document}